\documentclass[aip,jap,amsmath,amssymb,reprint]{revtex4-1}

\usepackage{graphicx}% Include figure files
\usepackage{dcolumn}% Align table columns on decimal point
\usepackage{bm}% bold math

\usepackage[utf8]{inputenc}
\usepackage[T1]{fontenc}
\usepackage{mathptmx}

\usepackage{upgreek} %upright greek symbols
\usepackage{xcolor} %colored text mainly for comments for Martin

\newcommand{\Mark}[1]{{\color{black}{#1}}}

\begin{document}

\title[HAR TES X-ray]{High Aspect Ratio Transition Edge Sensors for X-ray Spectrometry}

\author{M. de Wit}
 \email{M.de.Wit@sron.nl}
\author{L. Gottardi}
\author{E. Taralli}
\author{K. Nagayoshi}
\author{M.L. Ridder} 
\author{H. Akamatsu}
\author{M.P. Bruijn}
\author{M. D'Andrea}
\author{J. van der Kuur}
\author{K. Ravensberg}
\author{D. Vaccaro}
\author{S. Visser}
 \affiliation{NWO-I/SRON Netherlands Institute for Space Research, Sorbonnelaan 2, 3584 CA Utrecht, The Netherlands} 
\author{J.R. Gao}
 \affiliation{NWO-I/SRON Netherlands Institute for Space Research, Sorbonnelaan 2, 3584 CA Utrecht, The Netherlands}
 \affiliation{Faculty of Applied Science, Delft University of Technology, Lorentzweg 1, 2628 CJ Delft, The Netherlands}
\author{J.-W.A. den Herder}
 \affiliation{NWO-I/SRON Netherlands Institute for Space Research, Sorbonnelaan 2, 3584 CA Utrecht, The Netherlands}

\date{\today}

\begin{abstract}
We are developing large TES arrays in combination with FDM readout for the next generation of X-ray space observatories. For operation under AC-bias, the TESs have to be carefully designed and optimized. In particular, the use of high aspect ratio devices will help to mitigate non-ideal behaviour due to the weak-link effect. In this paper, we present a full characterization of a TES array containing five different device geometries, with aspect ratios (width:length) ranging from 1:2 up to 1:6. The complex impedance of all geometries is measured in different bias configurations to study the evolution of the small-signal limit superconducting transition parameters $\alpha$ and $\beta$, as well as the excess noise. We show that high aspect ratio devices with properly tuned critical temperatures (around 90 mK) can achieve excellent energy resolution, \Mark{with an array average of 2.03~$\pm$~0.17~eV at 5.9~keV and a best achieved resolution of 1.63~$\pm$~0.17~eV.} This demonstrates that AC-biased TESs can achieve a very competitive performance compared to DC-biased TESs. The results have motivated a push to even more extreme device geometries currently in development.
\end{abstract}

\maketitle

\section{Introduction} \label{sec:intro}

\Mark{We are} developing large arrays of Ti/Au Transition Edge Sensors (TESs) optimized to be readout using Frequency Domain Multiplexing (FDM), intended for the next generation of X-ray observatories. One of such observatories will be ESA's Advanced Telescope for High Energy Astrophysics (Athena), scheduled for launch in the early 2030's. This mission will study the hot and energetic universe by detecting X-ray emission or absorption near compact objects including Active Galactic Nuclei (AGN) and defuse emission from extended sources ranging from SuperNova Remnants (SNR) to clusters of galaxies\cite{Barret2016,Barret2020}. One of the two instruments on board will be the X-ray Integral Field Unit (X-IFU), a detector consisting of an array of over 3000 Transition Edge Sensor (TES) calorimeters, each with an intrinsic energy resolution of 2.0~eV for 7~keV X-rays \cite{Barret2016,Pajot2018,DenHartog2018}. The SRON TES array has been selected as the backup option for the Athena mission. Furthermore, the Ti/Au TES detectors are one of the candidate detectors for the Hot Universe Baryon Surveyor, HUBS \cite{Cui2020}.

TESs can be used as imaging detectors with a very high spectral resolution, with applications over a wide range of wavelengths, from sub-millimeter waves to gamma-rays \cite{Niwa2017,Khosropanah2016,Smith2012,Horansky2007,Bandler2013}. In essence, a TES calorimeter consists of three main components: an absorber, a thermometer, and a weak link to a thermal bath. An absorber is used to absorb the photon energy and convert it into heat. This heat is transferred to a thin superconductor, biased to within the superconducting transition, that acts as a very sensitive thermometer. This sensor is then thermally isolated from the bath, typically by fabricating it on a silicon nitride membrane. The thermal and spectral bandwidths of the calorimeter are determined by the combined properties of these three main components. Excellent reviews have been written about both the physics and applications of TESs \cite{Irwin2005,Ullom2015}. In terms of the efficient readout of many TES devices, the two main techniques are Time Division Multiplexing (TDM), in which the TESs are DC-biased, and FDM, in which the TESs are AC-biased. Additional techniques such as Code Division Multiplexing (CDM)\cite{Morgan2016} and microwave SQUID multiplexing\cite{Bennett2019} are under development.

Traditionally, TESs used for X-ray calorimetry have been pushed to low resistance values \Mark{($R_n \lesssim 10~\mathrm{m\Omega}$)}. This type of device is very suitable for the DC-biased TDM readout since they typically have very sharp superconducting transitions and low internal thermal fluctuation noise between different parts of the calorimeter \cite{Wakeham2019}, leading to very good energy resolution \cite{Smith2012}. However, these low-Ohmic devices are generally unsuited for AC-biased readout. First, low-ohmic TESs suffer severely from AC losses in the normal metal structures surrounding the TES, as the losses can be a significant fraction of the TES resistance \cite{Gottardi2017,Sakai2018}. \Mark{Secondly, short, low-Ohmic TESs are in general more affected by the weak-link effect, in which (a part of) the bilayer is proximized by the presence of the niobium leads. This can result in changes in the effective critical temperature ($T_C$) of the bilayer\cite{Ridder2020}, a fraunhofer-like dependence of the critical current on the magnetic field\cite{Gottardi2014}, and steps in the superconducting transition \cite{Gottardi2018}.}

\Mark{To understand why low-Ohmic devices are more sensitive to the weak-link effect, we quickly review the main results reported by Gottardi \textit{et al.}\cite{Gottardi2018}, who compared the weak-link behaviour of several TES-based microcalorimeters and bolometers. It is stated that the amplitude of the Josephson current decreases with the increase of the superconducting phase difference across the TES:
\begin{equation}
\phi = \frac{2\pi}{\Phi_0} \frac{V}{\omega} = \frac{2\pi}{\Phi_0} \frac{\sqrt{P R}}{\omega},
\end{equation}
where $P$ is the detector power, $R$ is the resistance, and $\omega$ the frequency of the AC-bias. From this it is clear that the resistance of the devices should be increased. However, at the same time it is required to go to low power devices to increase the multiplexing number of the FDM readout, increasing the weak-link effect. This further drives the demand for high resistance devices.}

\Mark{For this reason we are developing TESs with high normal resistances ($R_n \gtrsim$~100~$\mathrm{m\Omega}$). This has to be done carefully though: Fabricating high normal resistance devices by changing the resistivity of the bilayer is not a viable approach, as it was shown by Wakeham \textit{et al.} that this would increase the internal thermal fluctuation noise (ITFN)\cite{Wakeham2019}. Therefore, we are keeping a constant sheet resistance, and instead develop TESs with high aspect ratios (width:length up to 1:6) with values for the normal resistance of $\sim$~50-150~$\mathrm{m\Omega}$. Whether the ITFN also increases for TESs of equal sheet resistance but increasing total resistance is not completely clear and has yet to be experimentally verified.}

We have previously reported on the first generation of high-aspect ratio devices, where we focused purely on the energy resolution and noise-equivalent power (NEP)\cite{Taralli2019}. However, the critical temperature of those devices was about 110 mK, 20 mK higher than the target value. In the current work, we have improved on these devices by fine-tuning $T_c$ to the desired 90 mK, resulting in reduced Johnson and phonon noise and thereby an improvement of the energy resolution by a factor of 1.34. Additionally, we have extensively investigated the small-signal limit parameters of the transition obtained from measurements of the TESs complex impedance, as well as the bias-dependent excess noise obtained from noise spectra. \Mark{We will use the data to look for trends in the performance of the difference designs. We will try to determine if we have reached a limit in the improvements that can be achieved by using high aspect ratios, or whether there might be more optimized geometries for better performance under an AC-biased readout scheme.}

This paper is structured as follows: In Sec.~\ref{sec:setup}, we explain the details of the measured TES array, and the setup that was used for all the measurements. In Sec.~\ref{sec:TES_char}, a full characterization of all devices is presented, including the thermal properties, small-signal parameters obtained from complex impedance measurement, and excess noise values. Finally, the obtained energy resolution at 5.9~keV is shown in Sec.~\ref{sec:energy}.

\section{Detectors and Measurement Setup} \label{sec:setup}

\subsection{TES array} \label{sec:array}

\begin{figure}
\centering
\includegraphics[width=\columnwidth]{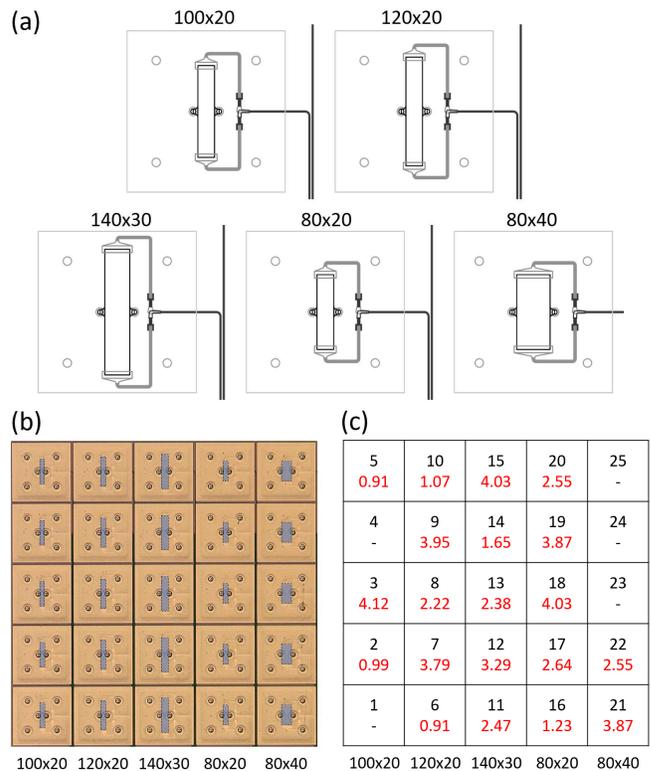} %{Figure_Setup/TES_Layout_2.pdf}
\caption{(a) The five different TES designs investigated in this report. The designs are all rectangular with dimensions 100$\times$20~$\upmu$m$^2$, 120$\times$20~$\upmu$m$^2$, 140$\times$30~$\upmu$m$^2$, 80$\times$20~$\upmu$m$^2$, and 80$\times$40~$\upmu$m$^2$. (b) Microscope image of the layout of the various TESs on the chip. Visible are the Au absorbers and the 6 stems connecting the absorber to the TESs (2 central stems) and to the SiN membrane (4 outer stems). The blue rectangles indicate the position and dimensions of the TESs underneath the absorber. (c) Schematic layout of the TES array. The top number (black) indicates the numbering of the TESs as used in the rest of this paper. The bottom number (red) gives the AC-bias frequency of each TES in units MHz. TESs 1, 4, 23, 24, and 25 have not been studied.}
\label{figure:TES}
\end{figure}

In this work, we have investigated 5 different TES designs, shown in Fig.~\ref{figure:TES}(a). The designs are rectangular with dimensions (length$\times$width) 100$\times$20~$\upmu$m$^2$, 120$\times$20~$\upmu$m$^2$, 140$\times$30~$\upmu$m$^2$, 80$\times$20~$\upmu$m$^2$, and 80$\times$40~$\upmu$m$^2$. This means we have designs with aspect ratios (width:length) ranging from 1:2 up to 1:6. The fabrication of the TES array is described in detail by Nagayoshi \textit{et al.} \cite{Nagayoshi2019}. Here we only summarize its most relevant properties.

The TESs are fabricated in a 5$\times$5 mixed array with a pitch size of 250~$\upmu$m, in which each column contains 5 identical designs to allow measurements on the various designs at different bias frequencies. The layout of the 5$\times$5 mixed array chip is shown in Fig.~\ref{figure:TES}(b), in which the positions of the TESs are indicated by the blue rectangles. The numbering of the TESs as used in the rest of this paper, as well as the AC-bias frequency at which they are measured can be read from Fig.~\ref{figure:TES}(c), where the black top number in each box indicates the TES number, and the red bottom number is the AC-bias frequency in units MHz.

The superconducting bilayer consist of Ti (35~nm) and Au (200~nm) fabricated on a 0.5~$\upmu$m thick SiN membrane used to thermally isolate the TES from the thermal bath. The bilayer is designed to have a critical temperature of $\sim$90~mK, and a squared normal resistance of 26~m$\Omega$/$\Box$. The bilayers are connected using niobium leads. Each TES is coupled to an 240$\times$240~$\upmu$m$^2$ absorber made of gold. In the final design for the X-IFU instrument, the absorber will also include bismuth, which has a high absorption efficiency (stopping power) whilst adding a negligible heat capacity. However, it was shown that bismuth absorbers can be responsible for low energy tails in the energy spectrum \cite{Yan2017,Gades2017}. Therefore, by using gold as an absorber we can better study the noise and physics intrinsic to the TES without additional complexity. The thickness of the gold absorber is 2.35~$\upmu$m, such that it has the same heat capacity as the final gold/bismuth absorber will have, approximately 0.85~pJ/K at 90~mK. The absorber is connected to the TES via two pillars in the center, and is further supported by four stems, one on each corner, on the SiN membrane. These supporting stems are visible in Fig.~\ref{figure:TES}(a) and (b).

This specific TES array has been measured and previously reported about by Taralli \textit{et al.} \cite{Taralli2019}. However, at that time the transition temperature of the TESs was about 110 mK, higher than the anticipated value of 90~mK, which limited the measured energy resolution to 2.4~-~2.8~eV. By baking the chip at 135$^{\circ}$C for 3 hours, a metal diffusion process has been used to reduce the transition temperature to the desired 90 mK \cite{VanDerHeijden2014, Nagayoshi2019}. This lower transition temperature has several beneficial effects, such as a reduction of the heat capacity as well as lower phonon and Johnson noise. The expected improvement of this action follows from the thermodynamic limit for the energy resolution of a calorimeter, given by $\langle \Delta E^2 \rangle = k_B T^2 C$, where $k_B$ is the Boltzmann constant, and $T$ and $C$ are the operating temperature and absorber heat capacity, respectively. Taking into account that $C$ scales linearly with the temperature\cite{Pobell2007}, we expect an improvement in the energy resolution of up to a factor of 1.35 compared to the previous result.

\subsection{FDM readout and cryogenic setup} \label{sec:FDM}

\begin{figure}
\centering
\includegraphics[width=\columnwidth]{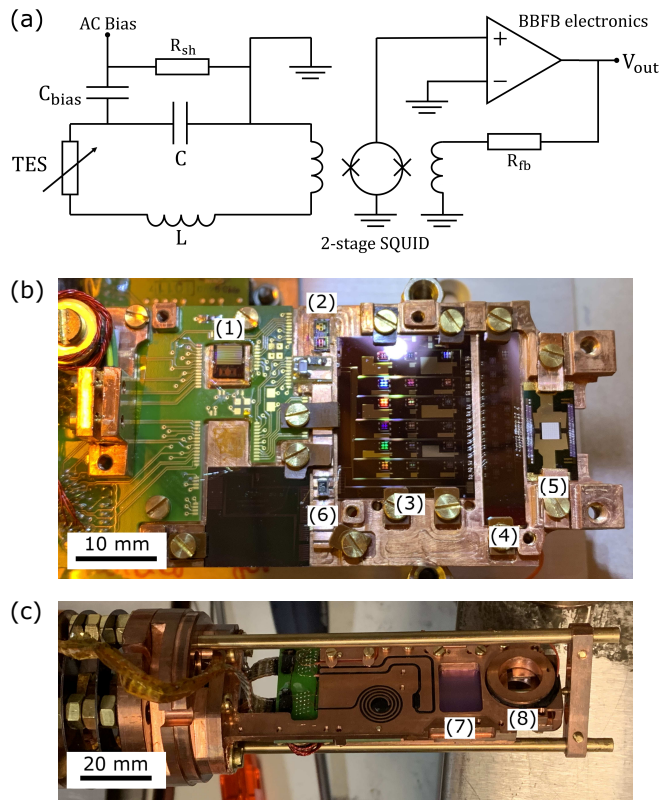} %{Figure_Setup/FDM_Setup_2.pdf}
\caption{(a) Schematic of the single-pixel AC-bias readout used to measure all TESs. Each TES is coupled to an LC resonator with a unique resonance frequency. The current that flows through the TES is measured using a two-stage SQUID developed by VTT. (b) Photograph of the copper bracket carrying all cryogenic components: (1) Amplifier SQUID; (2) front-end SQUID; (3) LC resonator chip; (4) interconnect chip; (5) TES array; (6) Shunt resistor. (c) Photograph of the backside of the setup shown in (b) mounted on the end of a Leiden-Cryogenics cold-insertable probe. A hole in the bracket is created at the location of the LC resonators (7) to reduce AC losses. A small Helmholtz coil (8) is used to apply perpendicular magnetic fields.}
\label{figure:FDM_setup}
\end{figure}

The characterization of the TES array described in the previous section is done by means of an FDM readout system developed at SRON. In this readout scheme, a TES is biased using a carrier signal with a frequency between 1~-~5~MHz. Each TES is coupled to a high-Q superconducting LC resonator to define a specific resonance frequency used for biasing \cite{Bruijn2012,Bruijn2018,Gottardi2019}. In this setup, all LC resonators have a coil inductance of 1~$\upmu$H. \Mark{While this inductance is too low to achieve critical damping necessary for good multiplexing performance, it is well-suited for studying the single pixel performance where one should be close to the small-signal limit depending on the detector normal resistance.} The current through each TES is measured using a two-stage SQUID assembly, developed by VTT \footnote{http://www.vttresearch.com/}. A schematic of this AC readout system is shown in Fig.~\ref{figure:FDM_setup}(a). The bias capacitor $C_{bias}$ is chosen such that $C/C_{bias} = 25$. Base-band feedback (BBFB) is used to increase the linearity and dynamic range of the SQUIDs \cite{DenHartog2009}. A more detailed explanation of the FDM readout is given by Akamatsu \textit{et al.} \cite{Akamatsu2014,Akamatsu2016,Akamatsu2020}. In order not to add additional complexity to the interpretation of the data, in this work the TESs are measured in single-pixel mode, where only one device is biased at a time, and all others are left in the superconducting state.

All cryogenic components of the FDM setup are mounted on a copper bracket, a picture of which can be seen in Fig.~\ref{figure:FDM_setup}(b). A copper collimator is placed over the array to prevent stray photons hitting the substrate. Fig.~\ref{figure:FDM_setup}(c) shows the backside of this bracket, including a Helmholtz coil that is placed over the TES array, and can be used to apply a homogeneous perpendicular magnetic field. Also visible is the backside of the LC resonator chip, where much of the copper is removed to reduce AC losses due to eddy currents. The setup is mounted at the bottom of a cold-insertable probe in a Leiden Cryogenics CF-400 dilution refrigerator \footnote{http://www.leidencryogenics.com/}. Magnetic shielding is achieved using a lead and cryoperm shield at the cold stage, and a mu-metal shield around the outside of the cryostat. The setup can be cooled to a base temperature of about 50 mK. All measurements in this paper were done at bath temperatures of $\sim$ 55 mK. A heater and germanium thermistor can be used to regulate the temperature to a stability within 1~$\upmu$K$_\mathrm{rms}$ at this bath temperature.

\section{TES Characterization} \label{sec:TES_char}

In order to compare the device performance with the theoretical models, in the following section we will characterize the relevant thermal and electrical properties of all pixels. First, we determine the critical temperature and thermal conductance of each TES. Then we evaluate the $\alpha$ and $\beta$ parameters from the measurement of the TES complex impedance. We subsequently use these parameters to fit the acquired noise spectra, in order to quantify the observed excess noise. Finally, for each device we estimate the expected energy resolution in the small-signal limit.

\subsection{Critical Temperature and Thermal Conductance}

\begin{figure}
\centering
\includegraphics[width=\columnwidth]{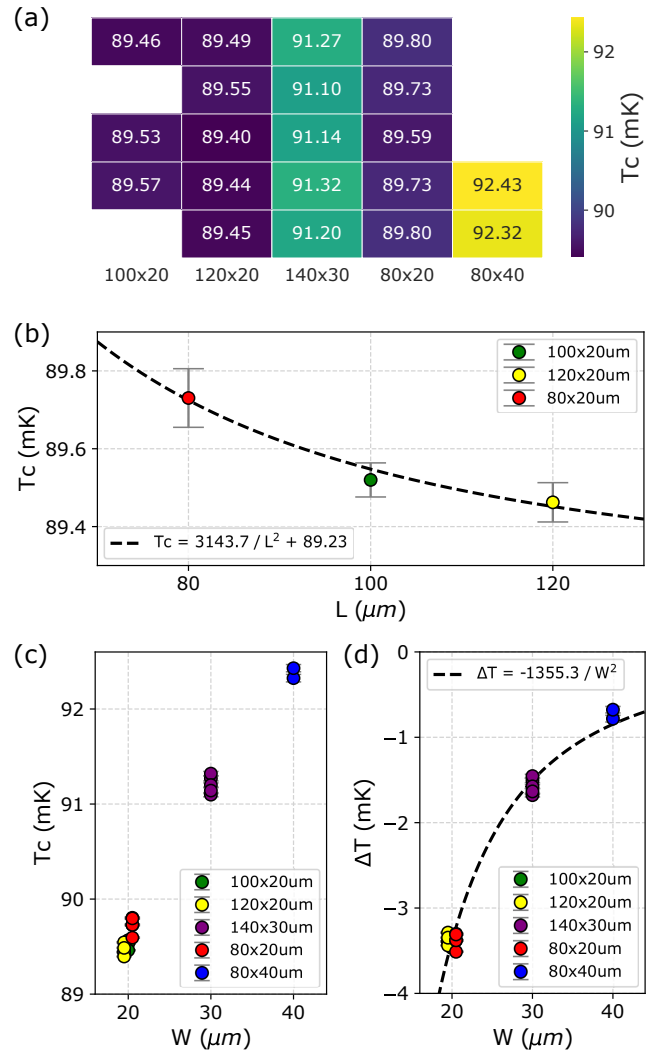} %{Figures_G_Tc/Figure_G_Tc_v3.pdf}
\caption{(a) Heatmap of the TES critical temperature $T_c$, clearly showing the dependence on geometry, but high uniformity within one geometry. (b) Averaged $T_c$ for all pixels with $W$~=~20~$\upmu$m as a function of TES length $L$. The error bar shows the standard deviation. The dashed black line is a fit of the data as explained in the main text. (c) $T_c$ versus TES width $W$. Note that the three pixel designs with $W$~=~20~$\upmu$m have been spaced for clarity. (d) Shift of the critical temperature $\Delta T$ due to the lateral inverse proximity effect, after correction for the longitudinal proximity effect. The dashed black line is a fit to the data using a $1/W^2$ scaling law.}
\label{figure:Tc}
\end{figure}

\begin{figure}
\centering
\includegraphics[width=\columnwidth]{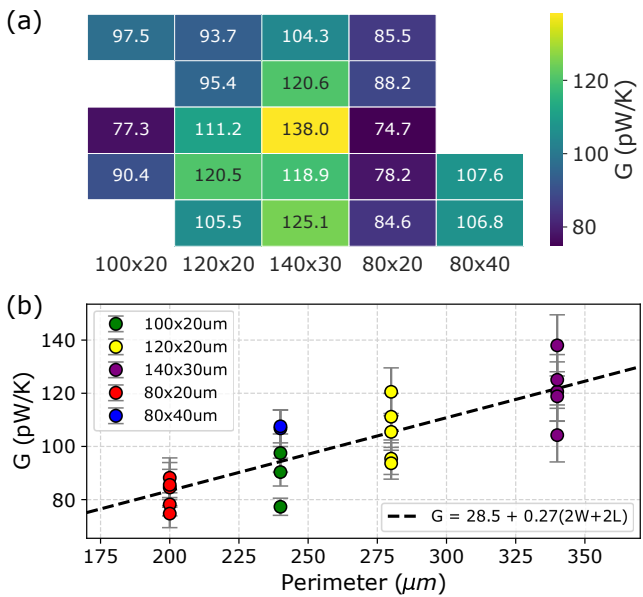} %{Figures_G_Tc/Figure_G_v4.pdf}
\caption{(a) Heatmap of the thermal conductance $G$ between the TES and thermal bath for the full TES array. (b) $G$ plotted as a function of the TES perimeter~=~$2W + 2L$. }
\label{figure:G}
\end{figure}

We have determined the thermal conductance and the critical temperature of all TESs in the array using measurements of IV-curves taken at different bath temperatures. From these IV curves, we extract the TES power (at $R = 0.5 R_n$) and construct power versus temperature (PT) curves. We fit these curves to a power law to extract values for the critical temperature and thermal conductance (at $T_c$). The obtained values are presented in Figs.~\ref{figure:Tc} and \ref{figure:G}.

The critical temperature, visible in Fig.~\ref{figure:Tc}(a), depends on both the width and the length of the devices. This dependence is due to a combination of the longitudinal proximity effect (LoPE) from the higher $T_c$ Nb leads connected to the bilayer and the lateral inverse proximity effect (LaiPE) from the overhanging Au edges of the bilayer. As a result, short devices show an increase in $T_c$, while narrow devices show a decrease. These effects are described in detail by Sadleir \textit{et al.} \cite{Sadleir2010,Sadleir2011}. In order to separate the contribution of the LoPE from that of the LaiPE, we first look at the TESs of equal width, but different lengths to identify the effect of the Nb leads. The results are then used to correct for the LoPE and look at the effect of the LaiPE in the devices of varying widths.

For the LoPE, the shift in $T_c$ scales with the spacing between the leads (equal to the length of the TES) following the relation \cite{Sadleir2010}:
\begin{equation}
\frac{\left( T_c - T_{c_0} \right)}{T_{c_0}} \propto 1/L^2
\end{equation}
with $T_{c_0}$ the transition temperature without the leads connected. The $1/L^2$ behavior has also been observed in other Ti/Au devices \cite{Ridder2020}. We can fit the data for the 100$\times$20~$\upmu$m$^2$, 120$\times$20~$\upmu$m$^2$, and 80$\times$20~$\upmu$m$^2$ pixels to this model, as shown in Fig.~\ref{figure:Tc}(b), where the dashed line indicates the best fit. We find a reasonable agreement with the predicted scaling, with $T_{c_0} = 89.2$~mK. Note that this value is lower than the intrinsic transition temperature of the Ti/Au bilayer due to the presence of the LaiPE from the bilayer edges. The fit confirms that the relation $T_c \propto 1/L^2$ remains valid even for these long devices.

The measured $T_c$ for all devices is shown in Fig~\ref{figure:Tc}(c) as a function of the width of the device. \Mark{As remarked previously, this is caused by the LaiPE originating from a small Au overhang. The overhang results from the HF etch used to pattern the titanium of the bilayer. From visual inspection it can be seen that the etching creates a Ti undercut of approximately 1~$\upmu$m at both sides of the bilayer, leaving an Au overhang. We expect that the thickness of the Au overhang is similar to that of the Au layer in the rest of the bilayer.} We check the validity of the LaiPE model by correcting the data of all geometries for the LoPE using the previous result. The corrected critical temperatures are fitted using a model that scales with 1/$W^2$, with $W$ the width of the device, as expected for the LaiPE \cite{Sadleir2010}. The results are shown in Fig.~\ref{figure:Tc}(d). \Mark{Shifts of the $T_c$ of up to 3.5~mK are observed for the smallest TES width of 20~$\upmu$m. This value is very similar to the values observed by Sadleir \textit{et al.}, who observed a shift of $\sim$~2-5~mK in devices with normal metal structures on top of the bilayer at the same 20~$\upmu$m spacing \cite{Sadleir2011}. This agreement supports our interpretation that the Au overhang is equivalent to a normal metal bank at the edge of the TES.}

Despite the large differences between geometries, the critical temperature is very constant within each design, with all identical devices within 0.2~mK of each other. \Mark{The small scatter in the $T_c$ results from a combination of the uniformity of the fabrication process and the measurement error.} We have measured similar scatter in $T_c$ for larger kilo-pixel arrays \footnote{\label{fn:note1}E. Taralli et al., submitted to RSI}. Additionally, because of the small values of the TES width the change in the transition temperature due to the LaiPE is predominant. This is something that should be considered when going to devices of even smaller width, in which case $T_c$ will be suppressed by several mK compared to the intrinsic value.

The thermal conductance, evaluated at the critical temperature, is shown in Figs. \ref{figure:G}(a) and (b). \Mark{From previous measurement of similar devices it is expected that the dominant process for the thermal conductance is two-dimensional radiative transport in the silicon nitride membrane\cite{Hoevers2005,Hays-Wehle2016}, in which case the thermal conductance scales with the TES perimeter. The current results are consistent with that interpretation. A comparison of the absolute value of our measured thermal conductance to earlier tests of similar devices fabricated by NASA Goddard reveal that our current results are about 20 pW/K above the expected values \cite{Kilbourne2007,Smith2014}.} This is the result of a parallel thermal conductance via the stems of the absorber. This additional thermal conductance can be reduced by using smaller diameters for the stems supporting the absorber on the membrane. The spread in the measured data for each geometry is most likely due to small deviations in the calibration of the front-end electronics. The power of the TESs depends on the geometry, ranging from 2~-~3~pW when biased at $R = 0.5 R_n$.

\subsection{Complex Impedance} \label{sec:ztes}

We have used measurements of the complex impedance to extract values for $\alpha$ and $\beta$, the unitless logarithmic derivatives of the resistance  versus temperature and versus current, respectively. Details about the method to measure the complex impedance under AC-bias can be found in Taralli \textit{et al.} \cite{Taralli2019a}. Briefly speaking, to measure the complex impedance, a small AC signal is added to the bias, and the resulting current through the TES is measured as a function of frequency. From this frequency-dependent response of the TES it is possible to extract the parameters that describe its thermal and electrical properties. This measurement is repeated at various points in the superconducting transition.

\begin{figure*}
\centering
\includegraphics[width=\linewidth]{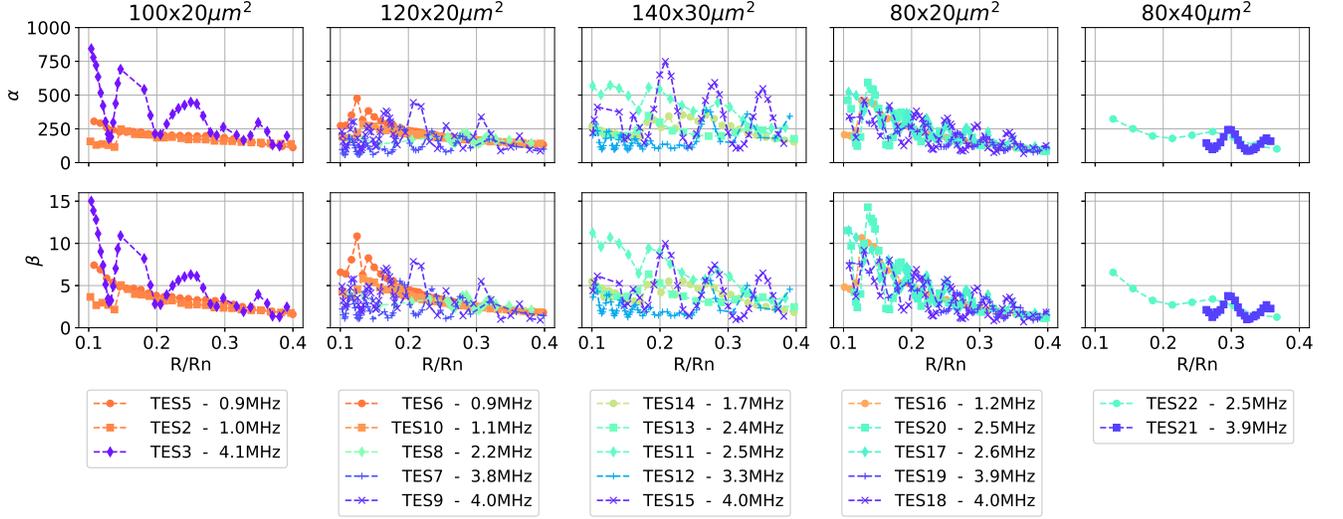} %{Figures_Complex_Impedance/alpha_beta_allgeoms}
\caption{$\alpha$ (top) and $\beta$ (bottom) for all geometries. The five columns show the data for the five different geometries, indicated at the top. Different colors in each figure correspond to different TESs within that geometry. Low frequency pixels show homogeneous transitions, high frequency pixels are affected by the weak-link effect, inducing the oscillations in the transition parameters.}
\label{figure:alphabeta}
\end{figure*}

We have fitted the complex impedance data with as a single thermal body including weak links effects, as outlined by Kozorezov \textit{et al.} \cite{Kozorezov2011}, which describes our devices quite well. The extracted $\alpha$ and $\beta$ for all geometries measured at $R/R_n$ between 10-40\% are shown in Fig.~\ref{figure:alphabeta}. The different colors in each figure show the different TESs measured at different bias frequencies. Since only 16 TESs can be measured per cryostat cycle, priority was given to high aspect ratio devices following our understanding of AC-biased TESs. For this reason, only two out of five 80$\times$40~$\upmu$m$^2$ devices were investigated, and relatively little data was collected for these devices.

All geometries show similar trends. The lower frequency pixels have a very smooth transition with little structure. In some TESs, an abrupt step is observed in both $\alpha$ and $\beta$, for instance visible in TES2. The high frequency pixels are characterized by strongly fluctuation values of $\alpha$ and $\beta$, especially for low values of $R/R_n$. The oscillations are caused by the weak-link effect \cite{Gottardi2014,Gottardi2018}. The low frequency pixels are less affected by this problem. All TESs for all geometries reach values of $\alpha \sim 300$ and $\beta \sim 5$, with the higher frequency pixels showing higher peak values due to the sharp steps in the transition. Note that $\alpha$ and $\beta$ are small-signal parameters, and when an X-ray is absorbed by a TES, typically a large part of the transition is sampled, smoothing these peaks.

\begin{figure*}
\centering
\includegraphics[width=\linewidth]{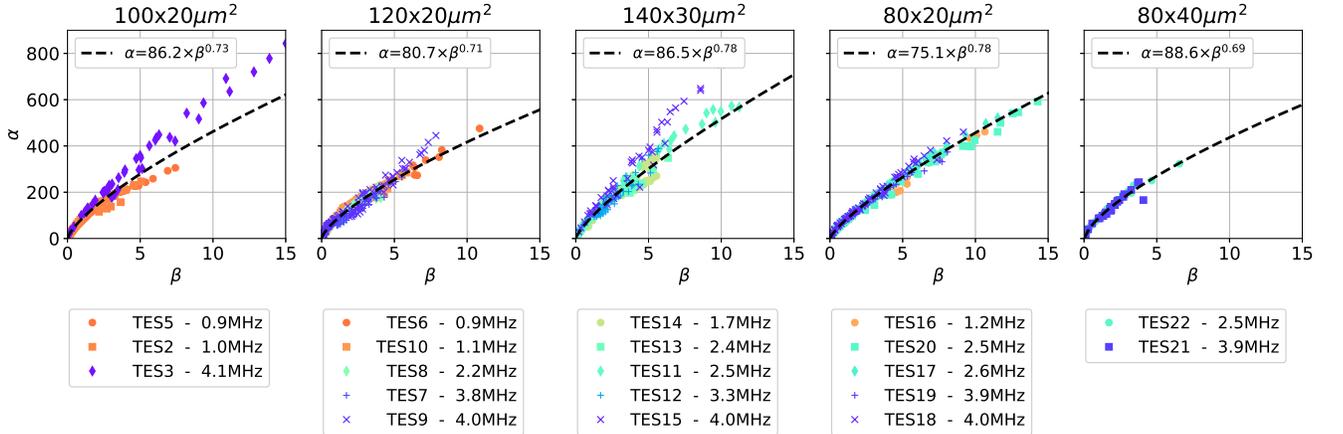} %{Figures_Complex_Impedance/alpha_vs_beta_fit_allgeoms.pdf}
\caption{Plot of the relation between $\alpha$ and $\beta$ for all geometries. Presented in this way, the results become nearly independent of the bias frequency. Dashed black lines are fits of the data to an empiric power law model, for data up to $\beta = 6$.}
\label{figure:alphabetafit}
\end{figure*}

The relation between $\alpha$ and $\beta$ is independent of the bias frequency, something which becomes clear when $\alpha$ is plotted versus $\beta$, as we have done in Fig.~\ref{figure:alphabetafit}. The TESs measured at different bias frequencies, which in Fig.~\ref{figure:alphabeta} show completely different behavior (i.e. the strong oscillations), now in general fall on a single line, with only a modest increase in the ratio between $\alpha$ and $\beta$ for the highest bias frequencies. The relation between $\alpha$ and $\beta$ follows a simple power law. This observation is shared for all geometries. The dashed black line is a fit of the data to the empirical function $\alpha = c \beta^{n}$. For all geometries we obtain similar values, with the pre-factor $c \sim 75-90$ and power $n \sim 2/3$.

\subsection{Excess noise} \label{sec:noise}

At each bias point where we measure the complex impedance we also measure noise spectra. The values obtained for $\alpha$ and $\beta$ from the complex impedance measurement are then used to fit these spectra. The main sources of noise in our TESs are phonon noise at low frequencies and TES Johnson noise at higher frequencies. Additionally, there is the white SQUID noise and shunt Johnson noise, but these terms are much smaller than the other factors. Details about this analysis are described by Taralli \textit{et al.}\cite{Taralli2019a}. We fit the high-frequency part of each noise spectrum to the following expression for the voltage noise:
\begin{equation}
V_n = \sqrt{4k_BTR (1+2\beta)(1+M^2)}
\end{equation}
Here $k_B$ is Boltzmann's constant and $M^2$ is the excess noise, given by the difference between the measured and predicted noise. This excess noise is typically explained by other noise sources, such as the weak-link effect described by the resistively shunted junction (RSJ) model \cite{Kozorezov2012,Gottardi2018}, the presence of phase-slip lines \cite{Bennett2012}, or ITFN \cite{Hoevers2000,Irwin2005,Takei2008,Maasilta2012,Wakeham2019}. In this paper, we will not go into a detailed analysis of the origin of the excess noise, but instead use it as a guideline to find the optimal TES design.

\begin{figure*}
\centering
\includegraphics[width=\linewidth]{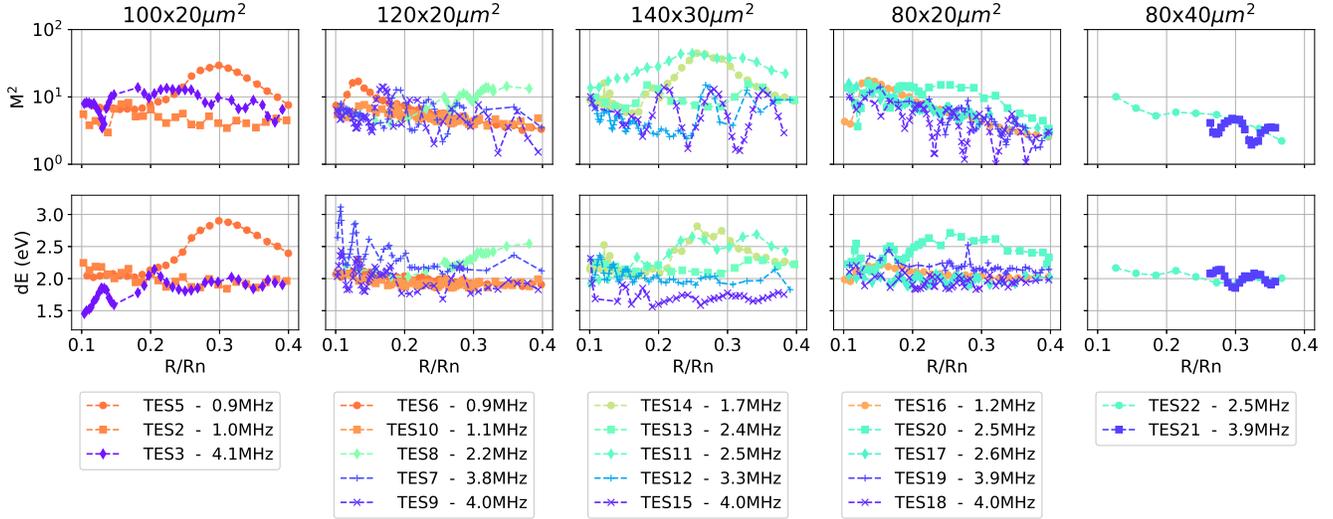} %{Figures_Complex_Impedance/M2_dE_allgeoms}
\caption{Excess Noise term $M^2$ and calculated energy resolution for all geometries. $M^2$ is correlated with $\alpha$, but the correlation is weak. All geometries have a similar small-signal limit energy resolution around 2 eV, but some high frequency pixels seem to be slightly better, as $\alpha$ increased faster than $M^2$.}
\label{figure:M2andFoM}
\end{figure*}

The obtained excess noise term $M^2$ is shown in Fig.~\ref{figure:M2andFoM} (top row). In general, the excess noise is correlated with $\alpha$ and $\beta$, with increasing oscillations for higher bias frequencies. However, there are some clear exceptions to this, where $M^2$ increases at $R/R_n > 20$\% (this can for example be seen clearly in the curve of TES5). At this moment it is unclear whether this additional bump in the excess noise is related to the readout, an intrinsic property of this TES design, or simply the result of individual bad devices. Excluding outliers, $M^2$ varies from 1~-~20 depending on the precise bias point. Several authors describe the existence of two regimes: $\alpha > 100$ and $\alpha< 100$. In the latter regime, they report that $M^2$ becomes less than one, so no significant excess noise\cite{Smith2013,Jethava2009}. We do not observe this relation.

The expected energy resolution of each device in the small-signal limit can be estimated using the following equation:
\begin{equation}
dE \approx 2.355 \sqrt{4k_{B}T_{0}^2 \cdot \frac{C}{\alpha} \sqrt{\left(1+2\beta \right) \left(1+M^2 \right)} }
\end{equation}
Here we assume that the total heat capacity $C$ is dominated by the absorber, and thus constant for all devices. The predicted energy resolution for all bias points for each device can be seen in the bottom row of Fig.~\ref{figure:M2andFoM}. In general, most of the devices for all geometries appear to have an expected resolution around 2.0~eV for a broad region of the superconducting transition. Despite the fact that the high frequency pixels show very inhomogeneous transitions, the expected energy resolution can be much lower than that of the TESs read  out at lower bias frequencies. In these cases, the high $M^2$ observed in some bias points is balanced by a high $\alpha$, with the ratio between these two scaling favorably for the higher bias frequencies. This balancing has been observed before in MoAu TESs under DC bias \cite{Miniussi2018}.

\begin{figure*}
\centering
\includegraphics[width=\linewidth]{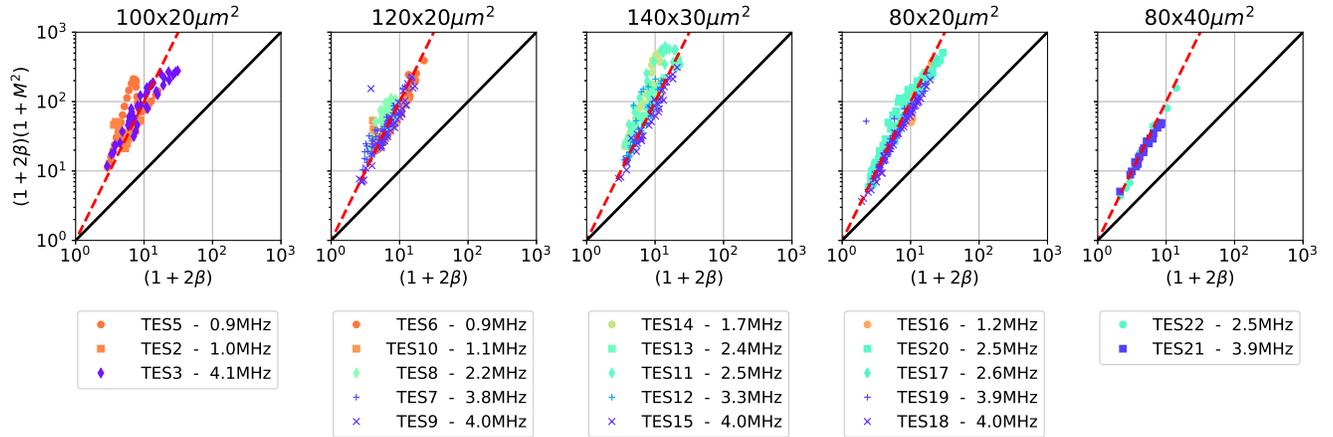} %{Figures_Complex_Impedance/M2_vs_beta_with_model.pdf}
\caption{Log-log plot showing the relation between $M^2$ and $\beta$. Solid black lines show the case when $M^2 = 0$. \Mark{The dashed red line is a guide to the eye to ease comparison of the different designs, with equal slope for all figures.}}
\label{figure:betaM2}
\end{figure*}

In Fig.~\ref{figure:betaM2} we show $(1+2\beta)(1+M^2)$ as a function of $(1+2\beta)$, the first order expansion term of the Johnson noise. Again, all frequencies and geometries exhibit common behavior. Note that at no point in the transition the data approaches the $M^2 = 0$ line, something which was reported in earlier reports on DC-biased TESs \cite{Smith2013}. \Mark{The deviation from the $M^2 = 0$ line is very similar for each geometry, clearly visible when looking at the dashed red line, which is a guide to the eye with equal slope for all figures.}
% dashed line assumes M2 = 2beta.

\section{X-ray Energy Resolution at 5.9 keV} \label{sec:energy}

\begin{figure}
\centering
\includegraphics[width=0.85\columnwidth]{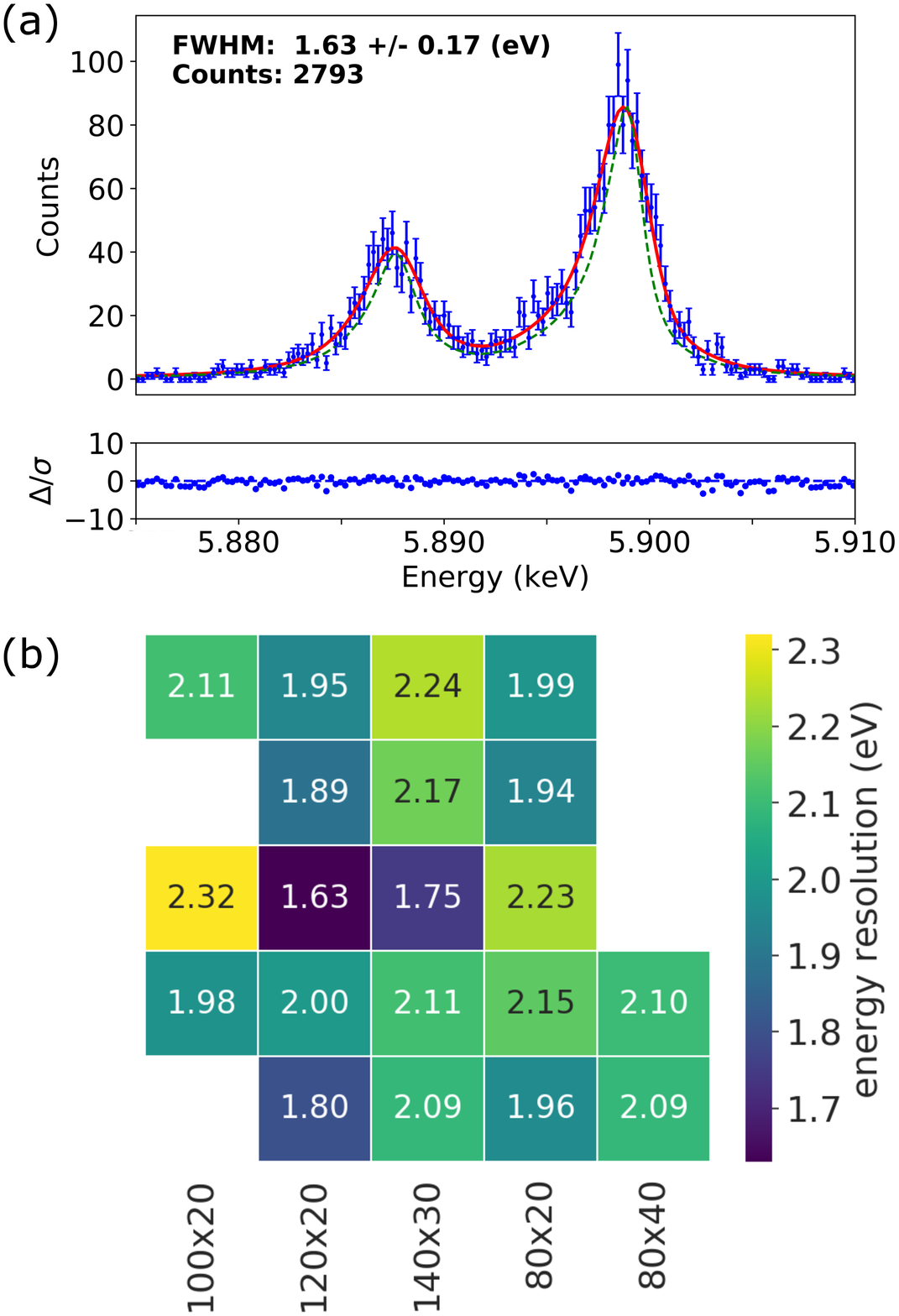} %{Figure_Energy/Figure_dE_V2.pdf}
\caption{(a) Measured energy spectrum of the Mn-K$\alpha$ complex from the $^{55}$Fe source, obtained from TES8 measured at 2.2~MHz, biased at $R/R_n = 22$\%. Blue dots indicate the measured spectrum, the solid red line the best-fit model. The green-dotted line is the natural line shape of the Mn-K$\alpha$ complex. Bottom panel shows the normalized residual. (b) Overview of the single-pixel best energy resolution for all TESs. Each value has an error bar of approximately $\pm$~0.17~eV.}
\label{figure:dE}
\end{figure}

In the previous section, we have calculated the expected small-signal energy resolution. However, due to non-linear effects in the TESs, the actual energy resolution can be far different. Therefore, we have measured the real energy resolution by exposing the TES array to a standard $^{55}$Fe source, providing Mn-K$\alpha$ X-rays at an energy of 5.9~keV and a count rate of approximately 1 count per second (cps) per pixel. Typically, for each spectrum we collect about 3000~-~4000 X-ray events to get a statistical error of about 0.15~-~0.18~eV for the reported energy resolution. An example of an X-ray energy spectrum is shown in Fig.~\ref{figure:dE}(a). This particular spectrum was measured using TES8, operated at a bias frequency of 2.2~MHz at $R/R_n = 22$\%.

The best energy resolution obtained for each TES can be seen in the heat map of Fig.~\ref{figure:dE}(b). The achieved energy resolution averaged over the entire array is 2.03~$\pm$~0.17~eV. This average energy resolution is a big improvement upon the 2.73~$\pm$~0.26~eV achieved on the same TES array before the reduction of the critical temperature \cite{Taralli2019}. The improvement by a factor of 1.34 is in almost perfect agreement with the expected improvement calculated in Sec. \ref{sec:setup}, indicating that our devices are indeed thermodynamically limited, and no significant second order effects are present. Only three of the TESs did not achieve an energy resolution below 2.2~eV \Mark{(ignoring the error bar)}. All of these devices were measured at the highest bias frequencies of 4.0 and 4.1~MHz, the frequencies that suffer most from the weak-link effect and AC losses. The best energy resolution was achieved for TES8 with the 120$\times$20~$\upmu$m$^2$ geometry, where we found a full width at half maximum of \Mark{1.63~$\pm$~0.17~eV}. This is the best resolution ever achieved with a Ti/Au TES, and within 0.1~eV from the best energy resolution reported for a single pixel under DC bias \cite{Smith2012,Bandler2013,Miniussi2018}.

\Mark{The energy resolutions shown in the heatmap in Fig.~\ref{figure:dE} were obtained after careful tuning of the pixel bias to optimize the performance. Changes in the bias current leads to different energy resolutions, as reflected by predicted energy resolution shown in Fig.~\ref{figure:M2andFoM}. Reducing the sensitivity of the pixels to the precise bias points is an import beneficial effect of reducing the weak-link effect. Alternatively, the consequences of the sensitivity could be mitigated by the use of a well-designed pixel tuning algorithm.}

\begin{figure}
\centering
\includegraphics[width=1\columnwidth]{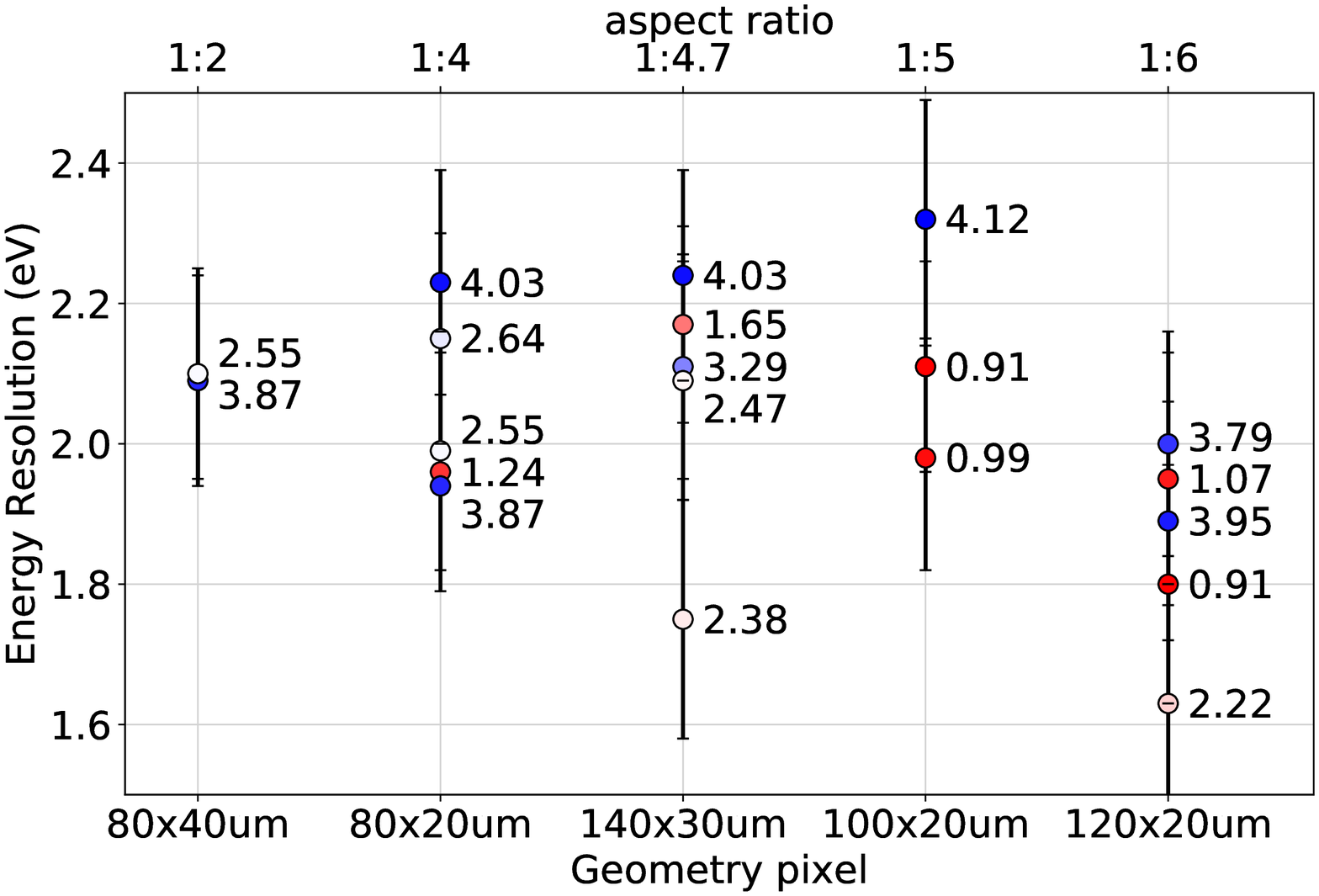} %{Figure_Energy/dEvsGeometry-allTESs.png}
\caption{Measured energy resolution for all geometries. The geometries are sorted to increasing aspect ratio. The number next to each point indicates the AC-bias frequency used to measure the TES. The color scale also indicates the bias frequency, where frequencies from low to high are coded in colors red to blue. \Mark{Each measurements is done at a bias point chosen to optimize the energy resolution.}}
\label{figure:dE_geometry}
\end{figure}

In Fig.~\ref{figure:dE_geometry}, we have plotted the measured X-ray resolution separated for the 5 different geometries. The number next to each data point indicates the bias frequency (in MHz) used to measure that pixel. From this figure, we can check the effect of the bias frequency at which a pixel is measured, and see which of the geometries outperforms the others. \Mark{The difference between the geometries is very small, with all geometries performing equal within the error.} Any apparent differences in performance between geometries are most likely not caused by the intrinsic noise of the pixels, which would have shown up in the complex impedance and noise measurements presented in Secs.~\ref{sec:ztes} and \ref{sec:noise}. Instead, they are a result of the homogeneity of the transition and the corresponding ease of finding the optimal bias point. Since the lower aspect ratio pixels suffer more from the weak-link effect, higher aspect ratios have wider tolerances in the operating parameters.

\Mark{No relation is observed between the bias frequency and the single pixel energy resolution. Both pixels measured at low bias frequencies as well as pixels measured at high bias frequencies achieve equal average energy resolution. This is clear after averaging the performance of the 6 lowest frequency pixels ($<$ 1.7~MHz), resulting in 2.00~$\pm$~0.12~eV, and the 6 highest frequency pixels ($>$ 3.7~MHz), leading to 2.11~$\pm$~0.16~eV, showing no statistically significant difference between the two sets.} Note that this observation is based on a small number of devices. A more detailed investigation of the bias frequency is reported by Taralli \textit{et al.}\footnotemark[3]. We \Mark{also} did not find a correlation between the optimal $R/R_n$ value and the bias frequency. The best energy resolutions for all devices are achieved for $R/R_n$ values ranging from 10-30\%.

Based on this data it is not possible to definitively say that higher aspect ratios lead to better X-ray performance. The data shows that, given that the devices are tuned to the right critical temperature, the AC-biased TESs perform at the same level as their DC-biased competitors. \Mark{In particular the 120$\times$20~$\upmu$m$^2$ devices are interesting, as these consistently achieved energy resolutions below \Mark{2.0~$\pm$~0.17~eV} for several bias points. These results show that even for a high aspect ratio and normal resistance there appears to be no degradation of the energy resolution nor an increase in the excess noise, a result that supports the idea that the ITFN indeed only scales with the resistivity and not with the total resistance. From this we expect that aspect ratios higher than 1:6 might also give very good results, possibly better than what we have seen here, should these devices really show a bigger reduction of the weak-link effect.}

\section{Conclusions}

We have investigated five different TES designs with varying aspect ratios ranging from 1:2 to 1:6. First, we have thermally characterized the devices. We have found that the critical temperature is dependent on the width and length of the devices, while the thermal conductance scales with the TES perimeter. All device geometries have similar small-signal parameters obtained from complex impedance measurements, and show comparable excess noise. The calculated energy resolution predicts that all devices are capable to measure 5.9~keV X-ray with a resolution below 2.0~eV \Mark{for a properly chosen bias point}.

All tested devices show excellent energy resolution at 5.9~keV, with an average of \Mark{2.03~$\pm$~0.17~eV} over the full array. The best performing pixel achieved a resolution \Mark{of 1.63~$\pm$~0.17~eV}, a record for an AC-biased TES X-ray calorimeter, as well as the first time such a resolution has been attained using a Ti/Au TES. This result is within 0.1~eV of the best performance achieved under DC-bias using devices fabricated at GSFC\cite{Miniussi2018}. \Mark{The achieved resolution is getting very close to 1.5 eV, the Athena energy resolution goal for X-rays with energies below 7~keV\cite{Pajot2018}.} That we so far have not reached the 1.5~eV level could in part be imputed to the AC-bias, but most likely a full understanding of the interplay between $\alpha$, $\beta$, and $M^2$ is required, something that is still under discussion for both AC- and DC-biased TESs.

Due to the small number of TESs per geometry that were investigated, together with the strong bias-frequency dependence of the TES behavior, more experiments are needed to reach hard conclusions on which geometry should be used. \Mark{However, it is clear that even for the highest aspect ratio devices we have studied here we see no limits in the performance, exemplified by the good overall results of the 120$\times$20~$\upmu$m$^2$ TESs. Given the idea that the weak-link effect should be reduced in high normal resistance devices, this opens the way to study even higher aspect ratio devices, in which we might find an optimum for the normal resistance and final performance.} New designs are currently under study.

\section{Acknowledgement}
This work is partly funded by European Space Agency (ESA) and coordinated with other European efforts under ESA CTP contract ITT AO/1-7947/14/NL/BW, and partly  by the European Union’s Horizon 2020 Program under the AHEAD (Activities for the High-Energy Astrophysics Domain) project with Grant Agreement Number 654215.

\section{Data Availability}
The data that support the findings of this study are available from the corresponding author upon reasonable request.

\bibliography{HAR_bibliography}

\end{document}